# Determination of step–edge barriers to interlayer transport from surface morphology during the initial stages of homoepitaxial growth

Pavel Šmilauer[*,†]

*Interdisciplinary Research Centre for Semiconductor Materials, The Blackett Laboratory, Imperial College, London SW7 2BZ, United Kingdom*

S. Harris[*]

*College of Engineering and Applied Sciences, SUNY, Stony Brook, NY 11794*

We use analytic formulae obtained from a simple model of crystal growth by molecular–beam epitaxy to determine step–edge barriers to interlayer transport. The method is based on information about the surface morphology at the onset of nucleation on top of first–layer islands in the submonolayer coverage regime of homoepitaxial growth. The formulae are tested using kinetic Monte Carlo simulations of a solid–on–solid model and applied to estimate step–edge barriers from scanning–tunneling microscopy data on initial stages of Fe(001), Pt(111), and Ag(111) homoepitaxy.

68.55.-a,05.70.Ln,68.35.Fx

Nearly thirty years ago, it was found in field–ion–microscopy experiments that interlayer transport on some metal surfaces is suppressed by additional activation barriers to hopping over step edges,[1] and such barriers were recently shown to be present on a semiconductor surface as well.[2] The consequences of this Ehrlich–Schwoebel (ES) barrier for growth on vicinal[3,4] and singular[5] surfaces have been theoretically investigated. In particular, study of the effects of the ES barrier for growth on a singular surface led to new insights into homoepitaxial growth modes[6] as well as into kinetic roughening of growing surfaces.[2] The ES barrier has emerged as a material parameter of an importance comparable to the surface diffusion barrier. It is, however, very difficult to determine this quantity experimentally.

Very recently, Meyer et al.[7] proposed (amongst other suggestions) a method of determining the ES barrier based on the surface morphology at the onset of nucleation on top of first–layer islands as seen by scanning tunneling microscopy (STM). In this paper, we provide a more consistent analytical treatment of this problem and derive a formula different from the one proposed by Meyer et al.. This formula and its modifications are tested using kinetic Monte Carlo (KMC) simulations and applied to STM data for Fe/Fe(001), Ag/Ag(111), and Pt/Pt(111) homoepitaxy.

We consider a model of the initial stages of homoepitaxy similar to the one proposed by Lewis and Anderson[8] with circular-shaped islands regularly distributed over a perfect singular surface (Fig. 1). Growth is initiated by a flux of atoms incoming to the surface. The adatom density $\rho$ on the surface increases until it reaches a critical value $\rho_c$ at a time $t=0$ when islands of radius $r_0 = 1$ separated by a distance $2L$ are formed. (Notice that all lengths are given in *units of the lattice constant* and are accordingly dimensionless.) We assume that the inter-island free–adatom density is then well approximated by

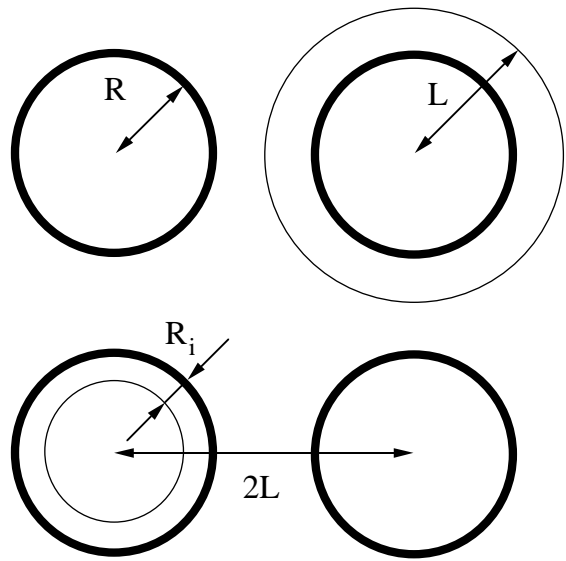

FIG. 1. Model geometry

its steady–state form with weakly time–dependent coefficients (quasi–static approximation[9]). We will consider relaxing the assumption of an instantaneous steady state later. Each island has an effective catchment area of radius $L$ with free–adatom flux equal to zero at a distance $r=L$. Based on these assumptions, the adatom density on the substrate immediately following nucleation is the solution of the diffusion equation

$$D\nabla^2 \rho + F = 0 \qquad (1)$$

with boundary conditions (cf. Ref. 9, Section 7.4) $d\rho/dr \,|_{r=R(t)} = \rho(R(t))$ (and thus also $d\rho/dr \,|_{r=r_0 \equiv R(0)} = \rho(r_0)$) and $d\rho/dr \,|_{r=L} = 0$, where $R(t)$ is the island radius, $F$ is the incoming flux, and $D = \nu_0 \exp(-E_D/k_B T)$ is the diffusion coefficient with $\nu_0$ being the attempt frequency, $E_D$ the surface diffusion barrier, $T$ the substrate temperature, and $k_B$ Boltzmann's constant. The solu-

tion is

$$\rho(r) = A + B \ln r - \frac{r^2}{4} \frac{F}{D} \quad (2)$$

where $A$ and $B$ are weakly time–dependent coefficients. Since $\rho(L) \approx \rho_c$ immediately following nucleation, evaluation of $A$ and $B$ leads to

$$\rho_c = \frac{F}{D} \left[ \frac{L^2}{2r_0} + \frac{r_0(r_0-2)}{4} + \frac{L^2}{2} \ln\left(\frac{L}{r_0}\right) - \frac{L^2}{4} \right] \quad (3)$$

The free–adatom density on top of the islands satisfies Eq. (1) with a boundary condition $D \, d\rho/dr \,|_{r=R(t)} = S\rho(R(t))$ where $S = \nu_0 \exp(-E_{\text{tot}}/k_B T)$ (we assume the same pre-factor $\nu_0$ for $D$ and $S$). $E_{\text{tot}} = E_D + E_B$ is the *total* barrier to move over the step edge, and $E_B$ is the *additional* barrier (the ES barrier) which might be positive, zero, or even negative depending on the values of $E_{\text{tot}}$ and $E_D$.[7,10]

The first–layer islands grow until they reach a critical radius, $R_c$, when second–layer free–adatom density reaches a critical value and nucleation on top of the first–layer islands occurs. Just prior to second–layer nucleation, $\rho(0) = \rho_c$ and $D \, d\rho/dr \,|_{r=R_c} = S\rho(R_c)$. In this case the logarithmic term does not appear in the solution of Eq. (1) and we have

$$\rho_c = \frac{F R_c}{2} \left( \frac{R_c}{2D} + \frac{1}{S} \right) \quad (4)$$

so that from Eqs. (3) and (4)

$$\frac{D}{S} \equiv \exp\left(\frac{E_B}{k_B T}\right) = \frac{L^2}{r_0 R_c} + \frac{r_0(r_0-2)}{2R_c} + \frac{L^2}{R_c} \left[\ln\left(\frac{L}{r_0}\right) - \frac{1}{2}\right] - \frac{R_c}{2}. \quad (5)$$

From this equation, the ES barrier $E_B$ can be determined.[11]

The above results are based on the assumption that immediately following first–layer nucleation, the inter-island free adatom density is given by its steady–state value, Eq. (2). This is unlikely and we now consider the consequences of relaxing this assumption. We expect that immediately following the nucleation, $\rho(r) \approx \rho_c$ for $r_0 < r < L$ except near $r = r_0$ where $\rho(r)$ is reduced. However, for a large $D/F$ ratio typical for available experiments,[10,12–14] the adatom density quickly reorganizes itself with only modest island growth due to attachment of adatoms from the region adjacent to the island. We thus replace $r_0 = 1$ in Eq. (5) by a value of $R(t)$ at the moment when the steady state is achieved. An estimate $r_0 = 1 + L/10$, certainly conservative given the dominating role of diffusion, is used below.

We may also consider an alternative estimate of $\rho_c$ to use in place of Eq. (3). If we use the order–of–magnitude expression[15] $\rho_c \approx L^2 F/D$ we obtain

$$\frac{D}{S} \approx \frac{2L^2}{R_c} - \frac{R_c}{2} \quad (6)$$

Note that $2L$ can be used instead of $L$ in Eq. (6) (so that $D/S = 8L^2/R_c - R_c/2$), cf. discussion in Ref. 15. The numerical estimates based on the Eqs. (5) and (6) are given in Tab. I. It is worthwhile to note that the critical density $\rho_c$ can in principle be determined experimentally by measuring the amount of material deposited before the onset of *first–layer* nucleation. It would be very useful to complement the observation of the onset of second–layer nucleation by such a measurement. Using Eq. (4), the ES barrier can be then estimated without any uncertainty connected with $r_0$.

To test our formulae, we employed simulations of a solid–on–solid KMC model in which no bulk vacancies or overhangs are allowed. Two processes are included in the model, the deposition of surface atoms and their migration with rates depending on their local environment (number of lateral nearest neighbors). In the basic version of the model, the hopping rates are given by an Arrhenius expression, $k(E,T) = \nu_0 \exp(-E/k_B T)$, where $E$ is the activation barrier to hopping comprised of a substrate contribution $E_S$ (always present) and a lateral nearest–neighbor contribution $nE_N$ proportional to $n$, the number of the same–layer nearest neighbors in the initial position.[16] This model has been modified to include the ES barrier by considering out-of-plane next–nearest neighbors of an adatom before and after a hop,[17] but for the purpose of the present investigations we used a simpler variant in which there is an additional contribution, $E_B$, to the hopping barrier $E$ for all inter-layer hops. The simulations were performed on $400 \times 400$ lattices with parameter values $E_S = 0.75$ eV, $E_N = 0.18$ eV, $E_B = 0.15$ eV, at a flux $F = 0.025$ monolayer(ML)/s, and the temperatures $T = 400$ K and $T = 450$ K.

When dealing with results of simulations (and also with STM images) and comparing them to model predictions, one important issue must be addressed: what are the proper quantities corresponding to the radius $R$ and the inter-island distance $2L$ of model islands for a collection of islands of non–ideal shapes, with a distribution of sizes and inter-island distances? After testing different possibilities, we estimated the inter–island distance $2L$ from the island density and the critical radius $R_c$ from the average island size. A double-check on the consistency of the values obtained is provided by the requirement that the resulting critical coverage of the first layer in the model given by $\Theta_c = (R_c/L)^2$ (note that an alternative $\Theta_c = \pi R_c^2/4L^2$ does not make an appreciable difference in the obtained values of the ES barrier) corresponds to the actual first–layer coverage at the onset of second–layer nucleation in a simulation or an experiment.[18]

The model parameters ($\Theta_c$, $R_c$, $L$) and the values of the ES barrier obtained from our simulations are summarized in Tab. I. The values of $E_B$ extracted from the snapshots of the surface morphology[18] are seen to underestimate (for $T = 400$ K) or overestimate (for $T = 450$ K) the value $E_B = 0.15$ eV actually used in simulations. Generally, the value of the ES barrier estimated from Eq. (5) increases with $\Theta_c$. At a low $\Theta_c$, small islands are responsible for the error, whereas at a high $\Theta_c$, the onset of coalescence leads to an overestimate of the barrier. However, we believe that the agreement (within 10–20%) is reasonable given the model simplifications and we did not attempt to compensate for these effects.

TABLE I. Estimates of model parameters and the Ehrlich–Schwoebel barrier (in eV) using different approximations for simulational and experimental data. The first four lines are results of simulations.

|  | $\Theta_c$ | $R_c$ | $L$ | $E_B{}^a$ | $E_B{}^b$ | $E_B{}^c$ |
|---|---|---|---|---|---|---|
| 400 K, $R_i=0$ | 0.20 | 3.2 | 7.2 | 0.11–0.13 | 0.12–0.17 | 0.06–0.12 |
| 400 K, $R_i=1^d$ | 0.30 | 4.3 | 7.8 | 0.10–0.12 | 0.11–0.16 | 0.05–0.11 |
| 450 K, $R_i=0$ | 0.45 | 12.8 | 19.1 | 0.15–0.18 | 0.15–0.21 | 0.08–0.15 |
| 450 K, $R_i=2^d$ | 0.50 | 14.3 | 20.2 | 0.15–0.18 | 0.15–0.21 | 0.08–0.15 |
| Fe(001)[12,13] | 0.7 | 5.1 | 6.1 | 0.05–0.07 | 0.06–0.10 | 0.00–0.06 |
| Ag(111)[10] | 0.55 | 451 | 608 | 0.18–0.22 | 0.19–0.23 | 0.13–0.19 |
| Pt(111)[14] | 0.3 | 53 | 97 | 0.21–0.25 | 0.21–0.27 | 0.15–0.21 |

<sup>a</sup>Eq. (5), $r_0 = 1 + L/10$ (first value), $r_0 = 1$ (second value).
<sup>b</sup>Eq. (6), $2L$ used instead of $L$ for the upper bound.
<sup>c</sup>Ref. 7, Eq. (8), $R_0 = L$ (first value), $R_0 = 2L$ (second value).
<sup>d</sup>$E_B{}^a \approx 0.12 - 0.14$ eV (400 K) or $\approx 0.16 - 0.19$ eV (450 K) if the effect of incorporation is taken into account in Eq. (5), see text.

We next look at experimental results. There is a bigger uncertainty in this case because an image taken exactly at the onset of second–layer nucleation (at a critical coverage $\Theta_c$) is usually not available. Fortunately, the value of the ES barrier is not extremely sensitive to an error in $\Theta_c$ and, therefore, we used the values provided by the authors of Refs. 7, 14 and 19. The best STM data available are for Fe/Fe(001) homoepitaxy.[12,13] We used the results for growth of Fe at 20°C and estimated the average inter-island distance $2L$ from the observed density of islands $\approx 8 \times 10^{12}$ cm$^{-2}$ (Ref. 12). The critical radius $R_c$ was estimated from the first–layer coverage[19] of 0.7 ML at a total coverage of 0.8 ML. The value of the ES barrier we obtained (Tab. I) is close to the estimates of $\approx 0.05$ eV based on comparisons between the evolution of the surface roughness for the same STM data and KMC simulations.[20] Our estimate may be too high because $\Theta_c$ is rather high. Other available experimental data are for Pt(111) from Ref. 14 and for Ag(111) from Refs. 7 and 10. We used the estimates for $L$ and $\Theta_c$ provided by the authors of these publications and our estimates of $R_c$ from the critical coverage and their STM pictures. The results are summarized in Tab. I.[21] The values of $\Theta_c$ suggest that the ES barrier for Pt(111) may in fact be even higher, and the one for Ag(111) lower than our results.

When extracting values for physical constants based on comparison with a simple model (analytical or KMC), it is always important to try to estimate how sensitive they are to the model simplifications. For real experimental systems, additional factors not considered in the model used here might play a role and it seems most of these lead to underestimating the ES barrier. Let us briefly discuss some of the possibilities:

(i) *A "leaking" ES barrier.* It has been proposed based on microscopic calculations that the barrier to descend might be lower[22] or higher[23] for rough steps or near kinks. Also, there are two types of steps (usually called A and B steps) on fcc(111) surfaces and it has been found in Refs. 22–24 that the barrier is much lower for B–type steps.[25] The value of the ES barrier obtained from a simple model is an effective value (averaged over the edges of all islands) and may be thus rather different from a barrier at a straight (e.g., A–type) step.

(ii) *Fractal islands.* If the island morphology is significantly different from the compact one we assumed, the model cannot be applied straightforwardly. Meyer et al.[7] suggested that in the case of fractal morphology the island radius can be replaced with the width of fractal arms but this is certainly a very crude approximation. In any case, the barrier obtained from the model as outlined above will underestimate the real barrier of fractal islands because of more frequent attempts by adatoms to hop down the steps. A recently suggested effect,[22] that an adatom can be trapped near a descending step, will also lead to more frequent attempts to hop down the step and thus a too low value of the barrier obtained from the model used here.

(iii) *Post–deposition non–diffusional incorporation.* In some cases, some of the atoms deposited near island edges can descend to a lower layer directly following their impact onto the surface.[22,26,27] This effect will again lead to a smaller barrier obtained from the model as compared to the real one, but the error will depend strongly on the average island size. We can provide a quantitative estimate by modifying the model such that all atoms deposited near island edges within a distance $R_i$ (in the region $R - R_i \leq r \leq R$, see Fig. 1) immediately descend to the lower layer instead of staying on top of the island. We then have to replace the flux $F$ on top of the island by $F'$, $F' = F(1 - 2R_i/R + R_i^2/R^2)$. This does not affect Eq. (3), but in Eq. (4), $F$ must be replaced by $F'$ so that all but the last term on the right–hand side of Eq. (5) are multiplied by the ratio $F/F'$. To test the effect of incorporation, we repeated our simulations with a different deposition rule: instead of simply dropping a particle onto a randomly selected site, we search for a site with the highest coordination *and at a lower layer* within a square region of a linear dimension of $2R_i + 1$ ($R_i = 1$ and 2 were used for the temperature 400 K and 450 K, respectively). The results given in Tab. I show that the incorporation leads to a change in the estimate of the barrier of the order of 0.02 eV (400 K) and 0.01 eV (450 K). Note however, that even incorporation over the distance of several lattice constants would not change appreciably the estimates of the ES barrier for Ag(111) and Pt(111) because large island sizes make this effect unimportant.

In conclusion, we have investigated a simple model to determine the value of the activation barrier to interlayer transport from information on the surface morphology in the initial stages of homoepitaxial growth. The model has been tested using kinetic Monte Carlo simulations of homoepitaxial growth and found to provide good estimates of the value of the step–edge barrier. This method of extracting the step-edge barrier from experimental data is straightforward and easy to use: the only information necessary is the island density and first–layer (or total) coverage close to the onset of nucleation on top of first–layer islands. Although we used scanning–tunneling microscopy data, high–resolution diffraction measurements can be used as well.[28] Various processes not accounted for in the simple model we use lead usually to underestimating the real value of the barrier.

We thank J.A. Stroscio and D.T. Pierce for permission to use their unpublished experimental data. P. Š. thanks to J. Vrijmoeth and J.A. Meyer for valuable discussions and for communicating their results prior to publication. The NATO Collaborative Research Grant CRG 931508, and the support of Imperial College and the Research Development Corporation of Japan are gratefully acknowledged.

* Also at: Department of Physics, Imperial College, London SW7 2BZ, United Kingdom.
† On leave from Institute of Physics, Cukrovarnická 10, 162 00 Praha, Czech Republic; present address: HLRZ, Forschungszentrum Jülich, 52425 Jülich, Germany. Electronic address: pavel@hlrserv.hlrz.kfa-juelich.de.

[1] G. Ehrlich and F.G. Hudda, J. Chem. Phys. **44**, 1039 (1966).
[2] M.D. Johnson, C. Orme, A.W. Hunt, D. Graff, J.L. Sudijono, L.M. Sander, and B.G. Orr, Phys. Rev. Lett. **72**, 116 (1994).
[3] R.L. Schwoebel and E.J. Shipsey, J. Appl. Phys. **37**, 3682 (1966); R.L. Schwoebel, J. Appl. Phys. **40**, 614 (1969).
[4] S.G. Bales and A. Zangwill, Phys. Rev. B **41**, 5500 (1990).
[5] J. Villain, J. Phys. I **1**, 19 (1991).
[6] R. Kunkel, B. Poelsema, L.K. Verheij, and G. Comsa, Phys. Rev. Lett. **65**, 733 (1990); B. Poelsema, R. Kunkel, N. Nagel, A.F. Becker, G. Rosenfeld, L.K. Verheij, and G. Comsa, Appl. Phys. A **53**, 369 (1991).
[7] J.A. Meyer, J. Vrijmoeth, H.A. van der Vegt, E. Vlieg, and R.J. Behm (unpublished).
[8] B. Lewis and J.C. Anderson, *Nucleation and Growth of Thin Films*, (Academic Press, New York, 1978), Sec. 3.IX.D.
[9] R. Ghez, *A Primer of Diffusion Problems*, (Wiley, New York, 1988).
[10] J. Vrijmoeth, H.A. van der Vegt, J.A. Meyer, E. Vlieg, and R.J. Behm, Phys. Rev. Lett. **72**, 3843 (1994).
[11] Meyer *et al.*[7] used Eq. (4) to determine the adatom density not only on top but also *between* the first–layer islands. However, there is no underlying physical model to support their method and, moreover, it leads to a major error in the value of the ES barrier obtained (cf. Tab. I).
[12] J.A. Stroscio, D.T. Pierce, and R.A. Dragoset, Phys. Rev. Lett. **70**, 3615 (1993); J.A. Stroscio and D.T. Pierce, Phys. Rev. B **49**, 8522 (1993).
[13] J.A. Stroscio, D.T. Pierce, M. Stiles, A. Zangwill, and L.M. Sander (unpublished).
[14] M. Bott, T. Michely, and G. Comsa, Surf. Sci. **272**, 161 (1992).
[15] J. Villain, A. Pimpinelli, and D. Wolf, Comments Cond. Matt. Phys. **16**, 1 (1992).
[16] S. Clarke and D.D. Vvedensky, Phys. Rev. Lett. **58**, 2235 (1987); D.D. Vvedensky, S. Clarke, K.J. Hugill, A.K. Myers–Beaghton, and M.R. Wilby, in *Kinetics of Ordering and Growth at Surfaces*, edited by M. Lagally (Plenum, New York, 1990), p. 297.
[17] P. Šmilauer, M.R. Wilby, and D.D. Vvedensky, Phys. Rev. B **47**, 4119 (1993).
[18] The onset of nucleation on top of first–layer islands is sharp for the idealized analytical model used here but continuous in simulations and experiments, and some empirical rule must be used to establish the value of the critical coverage. In the calculations presented in Tab. I, the onset of second–layer nucleation was supposed to occur when approximately 1/10 of first-layer islands have islands composed of two or more atoms on top of them. In Ref. 7, the authors required approximately half of islands to have stable nuclei on top of them [J.A. Meyer, personal communication]. However, the values of parameters $R_c$ and $L$ used in Ref. 7 correspond in fact to much lower coverages than the authors indicated.
[19] J.A. Stroscio and D.T. Pierce (unpublished).
[20] J.G. Amar and F. Family (unpublished); M.C. Bartelt and J.W. Evans (unpublished).
[21] Fig. 3b in Ref. 14 shows the onset of the second–layer nucleation at the total coverage of 0.8±0.1 for Pt(111) growth at 628 K when most of the islands are already coalescing and form a percolating structure with a rather complicated shape. Therefore, the model we are using cannot be applied in a straightforward manner. If we nevertheless use it, we obtain values of $E_B$ in the range 0.3 − 0.4 eV which is definitely too high, in particular given the fact that island edges in the high temperature range are of a B–type (cf. Ref. 25). The fact that the barrier is overestimated at this very high critical coverage is in agreement with the trend seen for the simulation data in Tab. I.
[22] M. Villarba and H. Jónsson, Phys. Rev. B **49**, 2208 (1994); M. Villarba and H. Jónsson, Surf. Sci. **317**, 15 (1994).
[23] Y. Li and A.E. DePristo, Surf. Sci. **319**, 141 (1994).
[24] R. Wang and K.A. Fichthorn, Surf. Sci. **301**, 253 (1994).
[25] The lower barrier to descend from a B–type step might play an important role in promoting layer–by–layer growth of Pt(111) above 450 K (Ref. 6). Whereas the islands below ≈450 K are bounded predominantly by A–type steps, the islands above ≈480 K are bounded predominantly by B–type steps [T. Michely, M. Hohage, M. Bott, and G. Comsa, Phys. Rev. Lett. **70**, 3943 (1993)] and therefore the ES barrier might be significantly lower at higher substrate temperatures.
[26] P. Stoltze and J. Nørskov, Phys. Rev. B **48**, 5607 (1993).
[27] J.W. Evans, D.E. Sanders, P.A. Thiel, and A.E. DePristo, Phys. Rev. B **41**, 5410 (1990); J.W. Evans, *ibid.*, **43**, 3897 (1991); M.C. Bartelt and J.W. Evans, Surf. Sci. **314**, L835 (1994).
[28] See, e.g., J. Wollschläger, J. Falta, and M. Henzler, Appl. Phys. A **50**, 57 (1990).